\documentclass[prl,aps,twocolumn,superscriptaddress]{revtex4-2}
\usepackage{graphicx}
\usepackage{dcolumn}
\usepackage{bm}
\usepackage{lipsum}
\usepackage{ragged2e}
\usepackage{braket}
\usepackage{bbm}
\usepackage{lipsum}
\usepackage{color}
\usepackage{xcolor}
\usepackage{amsfonts}
\usepackage{amsmath}
\usepackage{amsmath,amssymb,latexsym}
\usepackage{wasysym,stmaryrd,marvosym}
\usepackage{mathrsfs}
\usepackage{soul}
\usepackage{dsfont}
\usepackage{float}

\usepackage[mathlines]{lineno}
\usepackage[colorinlistoftodos]{todonotes}
\usepackage[colorlinks=true, allcolors=blue]{hyperref}

\begin{document}

\newcommand{\ohio}{Department of Physics and Astronomy and Nanoscale and Quantum Phenomena Institute, Ohio University, Athens, Ohio 45701}

\title{Tuning electronic pairing by uniaxial strain in kagome lattices}

\author{M. A. Mojarro}
\email{mm232521@ohio.edu}
\author{Sergio E. Ulloa}
\affiliation{\ohio}

\date{\today}

\begin{abstract}
We study the interplay of attractive electron interactions and topological states in strained kagome lattices with spin-orbit coupling via a Hubbard Hamiltonian in the mean-field approximation.
In the unstrained lattice, there is a topological phase transition from a quantum spin Hall state to a charge density wave (CDW) with increasing interaction strength. 
Upon applying a uniform uniaxial strain to the lattice, we find a new phase with coexisting CDWs and topological states. For increasing interaction strength or strain, the system is driven into a pure CDW, signaling topological phase transitions.
The directionality (nematicity) of the CDW is controlled by the direction of the applied strain.
When $s$-wave electronic pairing is allowed, the system develops a superconducting order beyond a threshold attraction, which is
totally suppressed by the onset of a CDW with increasing interaction. 
Most interestingly, moderate strain allows the coexistence of superconductivity and CDWs for a range of interaction values.
This illustrates how electronic interactions and single-particle topological structures compete to create unusual correlated phases in kagome systems.
\end{abstract}


\maketitle


Geometric frustration, topological states, and rich correlated-electron phenomena have been widely recognized in kagome lattices, arising from unique crystal symmetries and corresponding band structure, featuring Dirac cones, van-Hove singularities, and a flat band \cite{Neupert2022}. 

The recently discovered family of superconducting compounds $A$V$_3$Sb$_5$ ($A$\,=\,K, Rb, Cs) \cite{ortiz2019,Neupert2022,Jiang2022}, with V atoms arranged in a kagome plane, has become a natural playground to investigate the long-sought unconventional states predicted in kagome lattices. 
Different topological surface states \cite{ortiz2020,Yong2022} and electronic orders have been observed in these materials, including unconventional 
charge density waves (CDWs) associated with Fermi surface nesting \cite{Mielke2022,Zhao2021,Jiang2021}, 
as well as electronic nematicity possibly intertwined with a superconducting order \cite{Nie2022}. 
However, other kagome compounds are found more suitable to study flat band physics effects, as the Fermi energy lies closer to that band in materials such as CoSn \cite{Kang2020,Chen2023,Cheng2023}, titanium-based compounds $A$Ti$_3$Bi$_5$ \cite{Yang2023,Jiang2023}, Ni$_3$In \cite{Ye2024}, and the recently discovered kagome metal CsCr$_3$Sb$_5$ \cite{liu2023}. 

The kagome Hubbard model has been extensively studied using diverse numerical methods at van Hove band filling $f=5/12$, close to where the Fermi energy often lies. 
The phase diagrams have been shown to exhibit a variety of correlated phases, such as charge, bond,
and spin orders, as well as unconventional pairing  \cite{Ferrari2022,Wu2021,Kiesel2012,kiesel2013,Wu2023,Yu2012,Wang2013,Lin2022,Denner2021}. 
Away from the van Hove filling, closer to the Dirac cones ($f=1/3$), tuning of on-site electron interaction strength leads to topological \cite{Titvinidze2022} and CDW \cite{Zhu2023} transitions, whereas longer range interactions may strongly modify the electronic orders \cite{Ferhat2014,Jing2023}.
When the Fermi level lies higher, in the vicinity of the flat band ($f=2/3$), the Hubbard model predicts a variety of spin-charge density waves and ferromagnetism \cite{Wen2010,lin2023}, as well as interaction-driven topological states from nearest-neighbor repulsion \cite{Liu2010soc,Sun2009,Wen2010}.

The symmetric nature of the lattice and resulting kagome single-particle spectrum suggests a strong tunability of their properties by external means. For instance, tuning of electronic instabilities by chemical doping of Ti \cite{Haitao2022}, Nb \cite{li2022,kato2022}, and Cr \cite{Saqlain2024} in CsV$_3$Sb$_5$ has shed light into the origin of correlated phenomena. Additionally, high sensitivity to strain on superconductivity \cite{Yin2021} and CDW orders \cite{Qian2021,wang2023,lin2024} in the V-based superconductors is beginning to be explored by both local and global probes. 
Tunability by strain in kagome magnets has been shown to control frustration in Y-kapellasite crystals, enhancing the Néel temperature \cite{wang2023-2}, while STM studies of epitaxial strain in FeSn films have suggested strong pseudo-magnetic fields \cite{Zhang2023}. 
Our work into the interplay between topology and electron correlations when strain is applied yields interesting insights into these phenomena. 

We study the tunability of CDWs by uniaxial strain in the kagome lattice with intrinsic spin-orbit coupling (SOC), as SOC gaps have been suggested in the V-  \cite{Golovanova2023,Gu2022} and Ti-based \cite{Yang2023,Jiang2023,Hu2023} kagome metals.
We consider the Hubbard model with attractive on-site electron interactions at different filling factors, allowing for the onset of superconducting pairing. 
Without pairing, strain results in the coexistence of CDWs and a quantum spin Hall state (QSHS) that sustains topological helical edge states.  As the magnitude of strain or interaction strength increases, a topological phase transition occurs to a pure CDW state with trivial topology. 
The direction of the applied strain determines the orientation of the charge ordering (nematic) phase. 
When superconducting pairing is considered, we choose a momentum-independent $s$-wave pairing potential, as a nodeless SC has been suggested in several kagome materials \cite{Mu2021,Duan2021,Gupta2022,Yin2021}. 
This superconducting pairing mixes (gaps) the helical edge states associated with SOC, while the strain allows the coexistence of SC and CDWs over a range of interaction strength that depends on the electron filling. 

\begin{figure}
    \centering
    \includegraphics[scale=0.24]{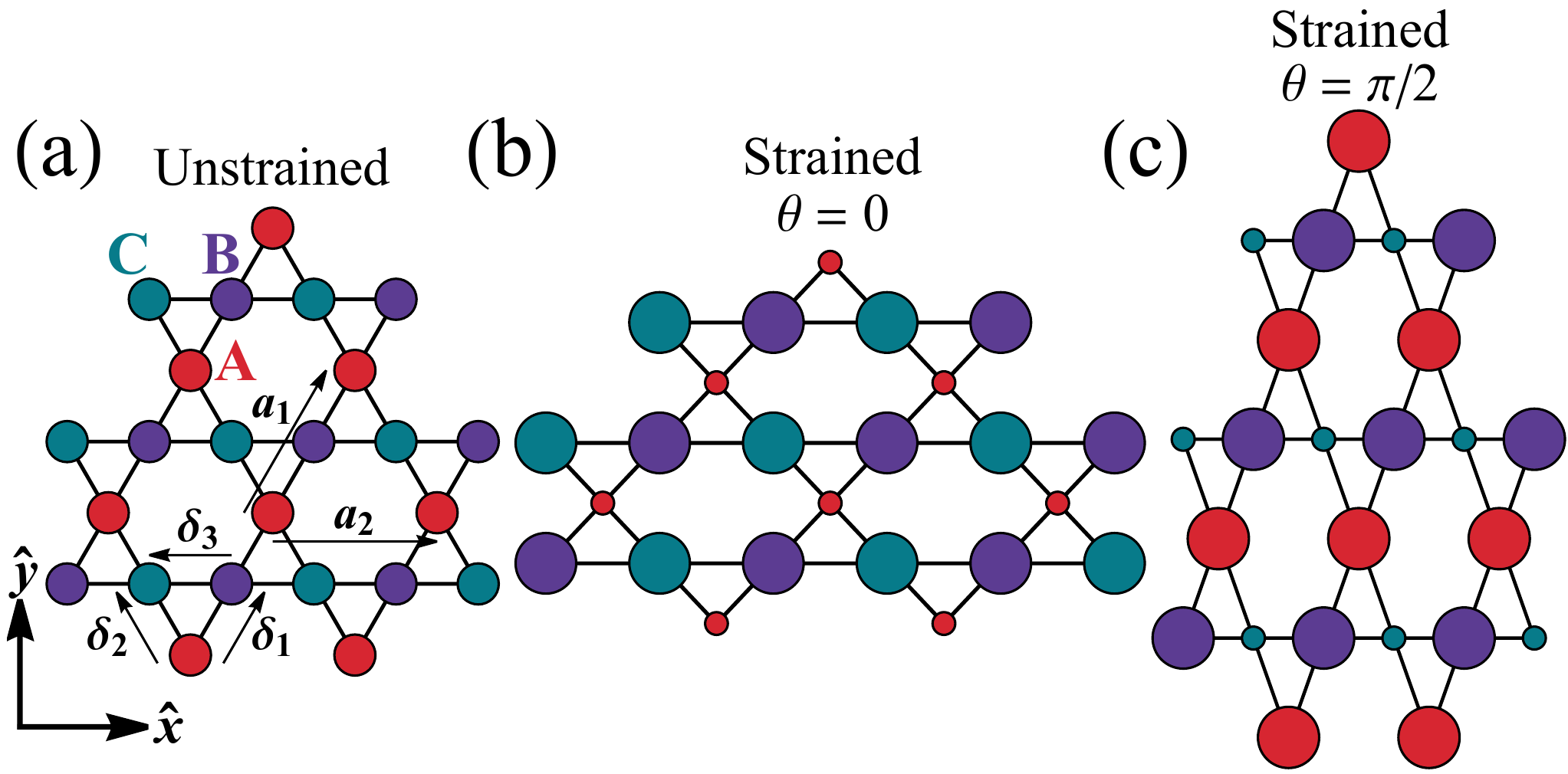}
    \caption{(a) Kagome lattice consisting of three triangular sublattices: A (red), B (purple), and C (cyan). The three nearest-neighbor vectors $\bm{\delta}_i$ and two primitive vectors $\bm{a}_i$ are shown. (b) and (c) show the CDW at $f=2/3$ of the strained lattice along the horizontal and vertical axis, respectively. Electron-rich (poor) sublattices are illustrated by large (small) sites. At $f=1/3$, electron-rich and electron-poor sites show the opposite occupancy.}
    \label{fig1}
\end{figure}

A uniform strain applied to a two-dimensional lattice transforms the atomic position vectors $\textbf{r}$ by a displacement field $\textbf{u}(\textbf{r})=\Bar{\epsilon}\cdot\textbf{r}$, where $\Bar{\epsilon}$ is the strain tensor given by
\begin{equation}
    \Bar{\epsilon}=
    \begin{pmatrix}
    \epsilon(\cos^2\theta-\nu\sin^2\theta) & \epsilon(1+\nu)\cos\theta\,\sin\theta\\
    \epsilon(1+\nu)\cos\theta\,\sin\theta & \epsilon(\sin^2\theta-\nu\cos^2\theta)
    \end{pmatrix}\,,
\end{equation}
with $\nu$ the Poisson ratio (taken as $\nu=0.165$ \cite{Androulidakis2018}), $\theta$ the direction of the strain with respect to the horizontal axis ($\bm{a}_2$), and $\epsilon$ the strain magnitude. This field leads to a set of lattice vectors $\bm{\delta}_i'=(\mathds{1}+\Bar{\epsilon})\cdot\bm{\delta}_i$ and $\bm{a}_i'=(\mathds{1}+\Bar{\epsilon})\cdot\bm{a}_i$, where $\mathds{1}$ is the $2\times2$ identity matrix, and the nearest-neighbor vectors of the unstrained tripartite kagome lattice are $\bm{\delta}_1=(1/2,\,\sqrt{3}/2)$, $\bm{\delta}_2=(-1/2,\,\sqrt{3}/2)$, $\bm{\delta}_3=\bm{\delta}_2-\bm{\delta}_1$, with primitive vectors $\bm{a}_1=2\,\bm{\delta}_1$, $\bm{a}_2=-2\,\bm{\delta}_3$ (in units of the inter-atomic distance), as shown in Fig.\ \ref{fig1}(a). 

We explore correlated and topological phases in the deformed kagome lattice by considering the attractive Hubbard model with intrinsic SOC,
\begin{equation}\label{ham}
    H=-\sum_{\braket{ij}\sigma}t_{ij}c_{i\sigma}^{\dagger}c_{j\sigma}-U\sum_{i}n_{i\uparrow}n_{i\downarrow}\pm\,i\lambda\sum_{\braket{ij}\sigma}\nu_{ij}c^{\dagger}_{i\sigma}c_{j\sigma}\,,
\end{equation}
where $c_{i\sigma}^{\dagger}$ ($c_{i\sigma}$) is the creation (annihilation) operator at site $i$ and spin $\sigma=\,\uparrow,\,\downarrow$, and $n_{i\sigma}=c_{i\sigma}^{\dagger}c_{i\sigma}$ is the spin-density operator. 
The symbol $\braket{ij}$ denotes sum over nearest-neighbor sites $i$ and $j$, $t_{ij}$ is the hopping energy, and $U>0$ is the on-site electron-electron attraction strength. In the last term, $\lambda$ is the intrinsic SOC strength (taken as $\lambda=0.08t$ throughout this work), the sign $+$ ($-$) corresponds to spin-up (down) electrons, and $\nu_{ij}=+1\,(-1)$ if the electron makes a clockwise (counterclockwise) turn when going from $i$ to $j$ through a neighboring site.
The effective attractive interaction introduced in the Hubbard model is associated with phonon-mediated interactions \cite{Balseiro1979}, although other mechanisms might also contribute.
The uniaxial strain makes the hopping energies strain-dependent as: $t_{ij}=t\,\text{exp}[-\beta(|\bm{\delta}_{ij}'|-1)]$, where $\bm{\delta}_{ij}'$ connects nearest-neighbor sites, $t$ is the hopping energy in the absence of strain, and $\beta$ is the Grüneisen parameter (taken as $\beta=3$ \cite{He2024}).

\begin{figure}
    \centering
    \includegraphics[scale=0.89]{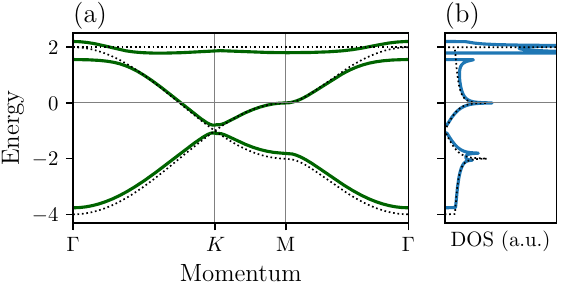}
    \caption{(a) Energy spectrum (in units of $t$) of a strained kagome lattice along $\theta=\pi/2$ with SOC (green) through a high-symmetry path of the Brillouin zone. The presence of SOC gaps the energy bands at both the $K$ and $\Gamma$ points. (b) Density of states (DOS) in blue shows gaps arising from the $K$ point, and a sharp structure from the $\Gamma$ mixings. Dotted lines correspond to the energy spectrum and DOS in the unstrained kagome lattice without SOC.}
    \label{fig2}
\end{figure}

The dynamics are well described in reciprocal space as the uniformly strained lattice remains periodic. 
This description reveals two graphene-like bands with Dirac cones and a nearly flat band at the top (see Fig.\,\ref{fig2}). The quadratic band-crossing point at the $\Gamma$ point of the Brillouin zone carries a Berry flux of $2\pi$ \cite{Sun2009,Montambaux2019} without strain, which splits into two Dirac points with Berry flux $\pi$ each when a strain is applied, reducing the $C_6$ symmetry of the lattice down to $C_2$ \cite{Mojarro2024}. 
The intrinsic SOC results in gaps of degenerate points in the energy spectrum as shown in Fig.\,\ref{fig2}, and drives the system into a QSHS protected by a $\mathds{Z}_2$ topological invariant \cite{Guo2009,Bolens2019} that supports helical edge states for $f=1/3$ and $f=2/3$. The topological phase is highly tunable by the strain as shown previously \cite{Mojarro2024}.

In order to include electronic interactions, we diagonalize Eq.\,\eqref{ham} employing a mean-field formalism, where the spin-density operator is approximated as $n_{i\sigma}\approx\braket{n_{i\sigma}}+\delta n_{i\sigma}$, where $\delta n_{i\sigma}$ is a small fluctuation around the mean density $\braket{n_{i\sigma}}$. Under this approximation, we decouple the on-site interactions as $n_{i\uparrow}n_{i\downarrow}\approx n_{i\uparrow}\braket{n_{i\downarrow}}+n_{i\downarrow}\braket{n_{i\uparrow}}-\braket{n_{i\uparrow}}\braket{n_{i\downarrow}}$, where quadratic terms in the fluctuations are neglected.

\begin{figure*}
    \centering
    \includegraphics[scale=0.8]{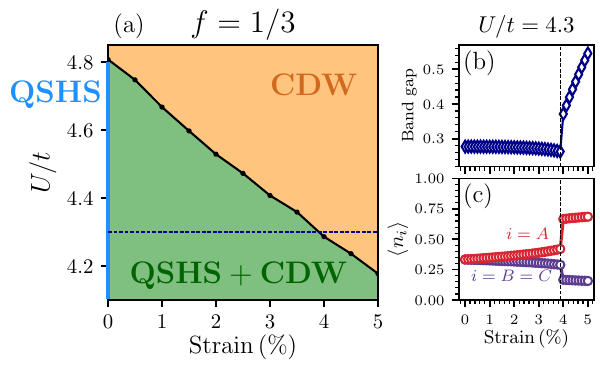}
    \includegraphics[scale=0.8]{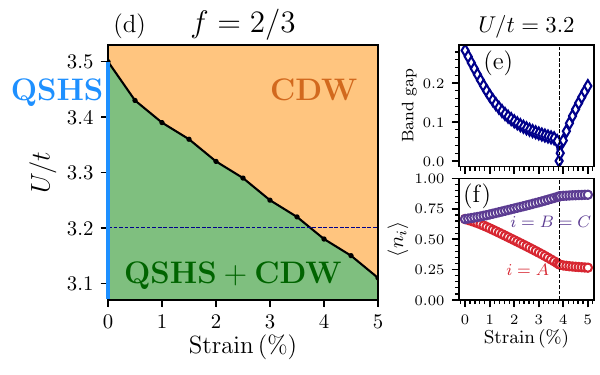}
    \caption{Mean-field phase diagrams as a function of attractive interaction $U$ and strain magnitude $\epsilon$ applied along the horizontal axis ($\theta=0$). The phase diagrams show a QSHS for vanishing strain (light blue), the coexistence of a QSHS and CDW (green), and a pure CDW state (orange).  
    (a) Phase diagram at $f=1/3$; band gap and order parameters $\braket{n_i}$ at $U/t=4.3$ (blue dotted line) shown in (b) and (c), respectively. Similarly, (d) shows the phase diagram at $f=2/3$, where band gap and order parameters at $U/t=3.2$ (blue dotted line) are shown in (e) and (f), respectively. Notice different $U/t$ scales in (a) and (d).}
    \label{fig3}
\end{figure*}

Motivated by the absence of local magnetic moments in the $A$V$_3$Sb$_5$ family \cite{ortiz2019,Kenney2021}, we set $\braket{n_{i\uparrow}}=\braket{n_{i\downarrow}}=\braket{n_i}$ ($i=\text{A,\,B,\,C}$). 
We allow the possibility of CDW formation described by two electron-poor (rich) sites and one electron-rich (poor) site within the unit cell following the triangle rule [see Figs.\,\ref{fig1}(b)-(c)], which preserves lattice translational symmetry  \cite{Nishimoto2010,Wen2010}, and results in lower overall energy in the attractive Hubbard model \cite{Zhu2023}.

In the mean-field approximation, Eq.\ \eqref{ham} can be written in reciprocal space as $H=\sum_{\bf k}\psi_{\bf k}^{\dagger}\left[H_{tb}({\bf k})+H_{\text{SOC}}({\bf k})+H_U\right]\psi_{\bf k}+NU\big(\braket{{n}_{\text{A}}}^2+\braket{{n}_{\text{B}}}^2+\braket{{n}_{\text{C}}}^2\big)/3$, where $N$ is the total number of sites and
\begin{eqnarray}
     H_{tb}({\bf k})&=&-2\sigma_0\otimes
    \begin{pmatrix}
    0 & t_1\cos {\bf k}'_1 & t_2\cos{\bf k}'_2\\
    t_1\cos{\bf k}'_1 & 0 & t_3\cos{\bf k}'_3 \\
     t_2\cos{\bf k}'_2& t_3\cos{\bf k}'_3 & 0
    \end{pmatrix}\,,\\
     H_{\text{SOC}}(\textbf{k})&=& 2\lambda    \sigma_z\otimes\begin{pmatrix}
    0 & i\cos{\bf k}'_1 & -i\cos{\bf k}'_2\\
     -i\cos{\bf k}'_1 & 0 & i\cos{\bf k}'_3 \\
     i\cos{\bf k}'_2& -i\cos{\bf k}'_3 & 0
    \end{pmatrix}\,,\\
    H_U&=&-U\sigma_0\otimes\begin{pmatrix}
    \braket{n_{\text{A}}} & 0 & 0  \\
    0 & \braket{n_{\text{B}}} & 0\\
    0 & 0 & \braket{n_{\text{C}}}  
    \end{pmatrix}\,,
\end{eqnarray}
in the basis $\psi_{{\bf k}}=\left(c_{\textbf{k}\uparrow},\,c_{\textbf{k}\downarrow}\right)^{\text{T}}$, where $c_{\textbf{k}\sigma}$ is a 3-component vector associated with the three kagome sublattices,
${\bf k}=(k_x,\,k_y)$ is the electron wave vector, $\sigma_0$ and $(\sigma_x,\,\sigma_y,\,\sigma_z)$ are the identity matrix and Pauli matrices acting on spin space, respectively, $t_{i}=\,\text{exp}[-\beta(|\bm{\delta}_{i}'|-1)]$, and ${\bf k}'_i={\bf k}\cdot\bm{\delta}'_i$.
Diagonalization of the Bloch Hamiltonian $H({\bf k})=H_{tb}({\bf k})+H_{\text{SOC}}({\bf k})+H_U$  leads to the mean-field energy spectrum $\varepsilon({\bf k})$. 
Minimizing the total energy per site $\varepsilon_0=(1/N)\sum_{{\bf k}}'\varepsilon({\bf k})+U\big(\braket{{n}_{\text{A}}}^2+\braket{{n}_{\text{B}}}^2+\braket{{n}_{\text{C}}}^2\big)/3$, where the prime in the sum indicates the set of states below the Fermi energy $\varepsilon_F$, $\{{\bf k}|\varepsilon({\bf k})<\varepsilon_F\}$, yields the ground state local occupations (``order parameters") $\braket{n_i}$, subjected to the condition for a given filling factor, $\sum_i\braket{n_i}=3f$.
Fig.\,\ref{fig3} shows results for $3f=1,2$, when the strain is applied along the horizontal axis ($\theta=0$). 
Panel (a) shows the phase diagram at $f=1/3$ as a function of $U$ and strain magnitude. In the absence of strain, there is a transition from a QSHS with gapped Dirac cones to a CDW state at $U/t\approx4.8$, consisting of two electron-poor sites, $\braket{n_{\text{B}}}=\braket{n_{\text{C}}}$, and one electron-rich site, $\braket{n_{\text{A}}}$, in the unit cell. 
As the strain is applied, a region appears in the phase diagram where the topological state coexists with a low-contrast CDW, and the gapped Dirac cones move with increasing strain magnitude towards one of the M points of the Brillouin zone located at $\pi/[\sqrt{3}(1-\nu\epsilon)]\hat{{\bf k}}_y$ \cite{Mojarro2024}. 
A transition to a pure (trivial) CDW state is achieved with increasing strain or interaction.
Fig.\,\ref{fig3}(b) shows a typical behavior of the band gap for $U/t=4.3$ as a function of strain. The dotted vertical line indicates the transition from the coexisting topological phase to the trivial CDW  gap. Note that the band gap does not vanish at the transition point, which could be an artifact of the mean-field analysis. 
In the topological phase, the band gap between Dirac cones remains almost strain-independent until the transition, while in the pure CDW state, the band gap now at the M point increases rapidly with strain as $|U(\braket{n_\text{A}}-\braket{n_\text{B}})-2\sqrt{t_3^2+\lambda^2}|$. 
For the same $U$, Fig.\ \ref{fig3}(c) shows the order parameters $\braket{n_i}$ as a function of $\epsilon$, revealing the coexistence of the weak CDW ($\braket{n_\text{A}}-\braket{n_\text{B}} \lesssim  0.13$) with the QSHS, and an enhancement of the polarization with increasing strain magnitude (reaching $\braket{n_\text{A}}-\braket{n_\text{B}} \approx 0.5$).

For higher filling, $f=2/3$, 
Fig.\,\ref{fig3}(d) shows the resulting phase diagram. In the absence of strain, there is also a transition from a QSHS to a CDW state at $U/t\approx3.5$. In this case, the CDW state has two electron-rich, $\braket{n_{\text{B}}}=\braket{n_{\text{C}}}$, and one electron-poor site, $\braket{n_{\text{A}}}$, in the unit cell [see Fig.\,\ref{fig1}(b)]. 
The QSHS develops a CDW, which is strongly enhanced with strain, as shown in Fig.\,\ref{fig3}(f) for $U/t=3.2$.
A transition to a pure CDW state with a trivial gap is achieved for larger strain magnitude or interaction strength. 
In contrast to $f=1/3$, in this case, the band gap is seen to close at the transition point at the M point, and then acquires a dependence on strain shown in Fig.\ \ref{fig3}(e), given by $|U(\braket{n_\text{B}}-\braket{n_\text{A}})-2\sqrt{t_3^2+\lambda^2}|$, while the CDW saturates to $\braket{n_\text{B}}-\braket{n_\text{A}} \approx 0.6$.

We should mention that in the absence of SOC in the system, a gapless CDW phase develops for small $U$ and non-zero strain at both $f=1/3$ and $2/3$, with Dirac fermion excitations. This CDW eventually becomes gapped for increasing interaction strength.

\begin{figure}
    \centering
    \includegraphics[scale=0.63]{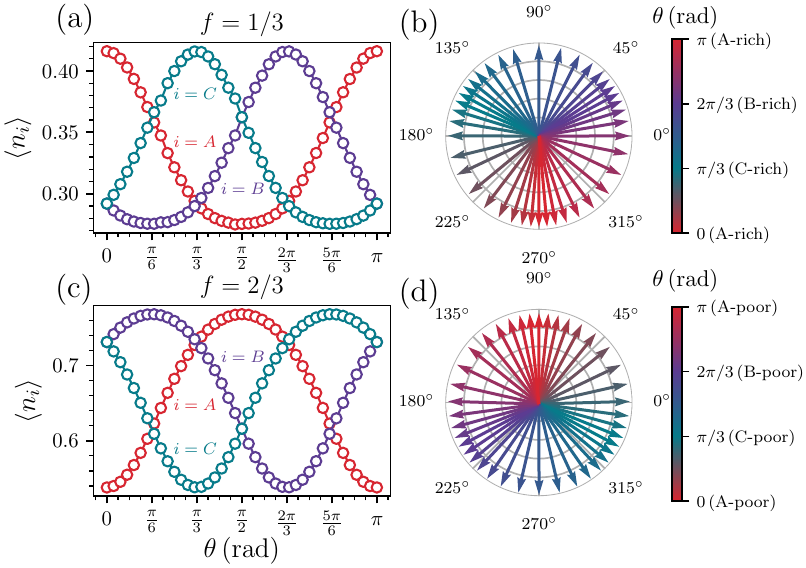}
    \caption{Strain direction dependence of the local charges at $f=1/3$ (with $U/t=4$) and $f=2/3$ (with $U/t=2.5$) shown in (a) and (c), respectively. The strain magnitude is $\epsilon=0.05$, such that the system at both fillings is in the coexisting QSHS with CDW (green region in Fig.\,\ref{fig3}). (b) and (d) show the direction of the nematicity vector $\hat{\textbf{e}}$ as the strain direction changes at $f=1/3,\,2/3$, respectively. The color of the $\hat{\textbf{e}}$ arrows indicates the electron occupancy in each sublattice, being rich or poor, as indicated by color bars. }
    \label{fig4}
\end{figure}

As mentioned, the CDW here does not break the original translational symmetry of the kagome lattice and it can be seen as a nematic phase in real space along the direction $\hat{\textbf{e}}=\textbf{Q}/|\textbf{Q}|$, where $\textbf{Q}=1/\sqrt{3}(\braket{n_{\text{B}}}-\braket{n_{\text{C}}})\hat{\textbf{x}}+1/3(\braket{n_{\text{B}}}+\braket{n_{\text{C}}}-2\braket{n_{\text{A}}})\hat{\textbf{y}}$ \cite{Sun2009,Wen2010}. 
We expect to see $\hat{\textbf{e}}$ vary as the strain direction is tuned. 
Fig.\,\ref{fig4} shows the sublattice charges and direction of nematicity as a function of the direction of strain for $\epsilon=0.05$ ($5\%$). 
The order parameters follow an oscillatory behavior at both $f=1/3$ and $2/3$ as a function of $\theta$, where electron-rich and electron-poor sublattices alternate. 
This oscillatory behavior results in the rotation of the nematicity vector as shown in Figs.\ \ref{fig4}(b)-(d) for $f=1/3$, $2/3$, respectively. 
Notice that when the strain is applied along the bonds of the kagome lattice, $
\theta=0,\,\pi/3,\,2\pi/3$, one of the sublattices becomes electron-rich (poor) while the other two sublattices are electron-poor (rich) at $f=1/3$ ($f=2/3$). In such cases, the direction of nematicity is perpendicular to the direction of the applied strain.  

\begin{figure}
    \centering
    \includegraphics[scale=0.65]{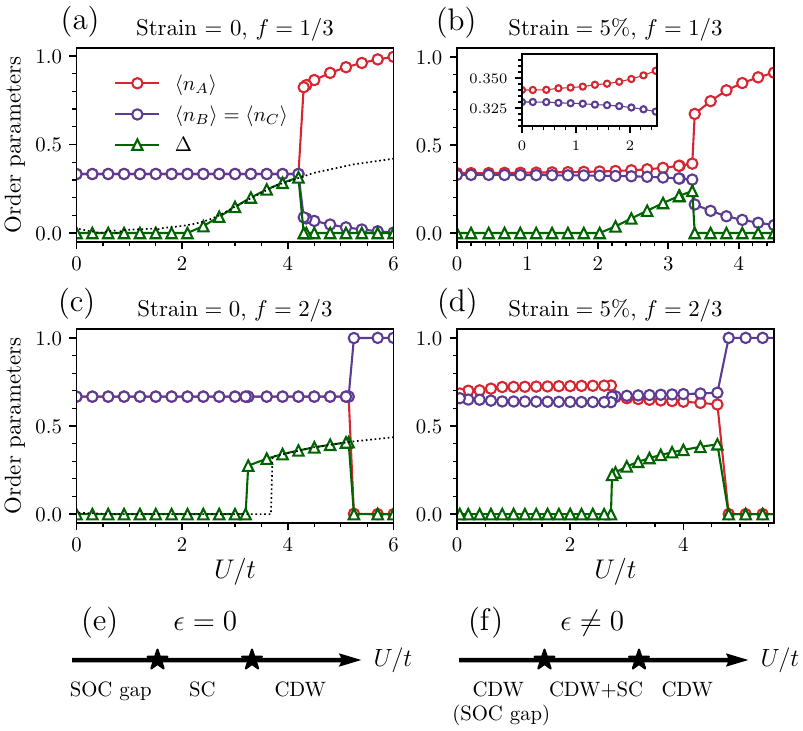}
    \includegraphics[scale=0.65]{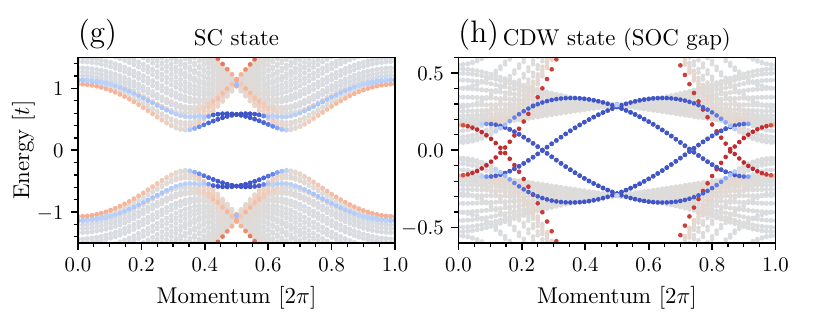}
    \caption{Mean-field order parameters $\braket{n_i}$ and $\Delta$ as a function of $U/t$ for (a) $f=1/3$ and (c) $f=2/3$ in the absence of strain, where a schematics of the phase diagram is shown in (e). The dotted lines are the superconducting order parameters when only pairing is considered. The initially gapless helical edge states due to SOC become fully gapped upon entering the SC state as seen in (g), where the energy spectrum of an infinite strip of kagome lattice with 15 unit cells in the finite direction is shown. Red (blue) represents the states in the top (bottom) of the strip (parameters at $f=1/3$, $U/t=3$ in (a)). 
    The cases for applied strain ($\epsilon=0.05$) in the horizontal direction are shown in (b) and (d) for $f=1/3$ and $f=2/3$, respectively. 
    A schematic phase diagram for the strain kagome lattice is shown in (f). In (h) are displayed the gapless helical edge states in the SOC gap when the system is in the CDW state before developing the SC order (parameters at $f=2/3$, $U/t=2.4$ in (d)).}
    \label{SCpairing}
\end{figure}

\paragraph{S-wave superconducting order parameter.}
To study the interplay between CDWs and superconductivity, and their tunability by strain, we include electronic pairing through the Bogoliubov–de Gennes (BdG) Hamiltonian, $H=\sum_{\bf k}\Psi_{\bf k}^\dagger H_{\text{BdG}}({\bf k})\Psi_{\bf k}+H_{U\Delta}$, where 
\begin{equation}\label{BdG}
    H_{\text{BdG}}=\begin{pmatrix}
        H({\bf k}) - \mu & \Gamma({\bf k})\\
        \Gamma^{\dagger}({\bf k}) &-H^*(-{\bf k})+\mu
    \end{pmatrix}\,,
\end{equation}
is written in the basis $\Psi_{{\bf k}}=\left(c_{\textbf{k}\uparrow},\,c_{\textbf{k}\downarrow},\,c_{-\textbf{k}\uparrow}^{\dagger},\,c_{-\textbf{k}\downarrow}^{\dagger}\right)^{\text{T}}$, with $\mu$ the chemical potential determined by the equation $f=1-\frac{d\varepsilon_0}{d\mu}$, and $\Gamma({\bf k})$ the pairing function \cite{zhao2006}.
We consider a momentum-independent $s$-wave pairing $\Gamma({\bf k})=-i\sigma_y\otimes\tau_0\,U\Delta $, where $\Delta=\braket{c_{i\downarrow}c_{i\uparrow}}$ is the local superconducting order parameter, and $\tau_0$ is the identity matrix in sublattice space. 
In this case, $H_{U\Delta}=NU\big(\braket{{n}_{\text{A}}}^2+\braket{{n}_{\text{B}}}^2+\braket{{n}_{\text{C}}}^2\big)/3+NU\Delta^2$.
The coupling between electrons and holes driven by $\Delta$ leads to an energy gap $2U\Delta$ in the well-known Bogoliubov quasiparticles spectrum $\xi_{\bf k}=\pm\sqrt{(\varepsilon({\bf k})-\mu)^2+U^2\Delta^2}$. Thus, whenever $\Delta\neq0$, the system develops a superconducting order.

The minimal total energy after diagonalization of Eq.\,\eqref{BdG} yields the mean-field order parameters $\braket{n_i}$ and $\Delta$, as shown in Fig.\,\ref{SCpairing}.
In the absence of strain, at $f=1/3$ the system develops a superconducting state with increasing gap in the region $2.1\lesssim U/t\lesssim 4.2$, which is totally suppressed by a fully gapped CDW order for $U/t\gtrsim4.2$ [see Fig.\,\ref{SCpairing}(a)]. 
Helical edge states due to SOC become gapped upon entering the superconducting state as time-reversal symmetry is preserved in this model.  
Fig.\,\ref{SCpairing}(g) shows such gapped edge states in an infinite stripe of kagome lattice in the superconducting state.
Time-reversal symmetry breaking due to a chiral flux phase may allow for gapless topological edge states in the presence of $s$-wave pairing, as proposed in the $A$V$_3$Sb$_5$ superconductors \cite{Gu2022}. 

Similar suppression of the superconducting order is observed at $f=2/3$ for large $U$. A superconducting gap manifests in the region $3.2\lesssim U/t\lesssim 5.2 $, 
whereas a pure CDW occurs for larger $U$, as shown in Fig.\ \ref{SCpairing}(c) for $\epsilon =0$. A schematic phase diagram of the unstrained kagome lattice is shown in Fig.\ \ref{SCpairing}(e).

The spatial anisotropy introduced by the uniaxial strain leads to the formation of a CDW within the superconducting state. This occurs in the regions $2\lesssim U/t\lesssim 3.4$ for $f=1/3$, and  $2.7 \lesssim U/t\lesssim 4.7$ 
for $f=2/3$, where the CDW order is somewhat weak [see Figs.\,\ref{SCpairing}(b) and (d)]. 
For increasing $U$, the superconducting order is suppressed by a strong CDW.
Our calculations also demonstrate that whenever $\epsilon\neq0$, before the onset of the superconducting gap and for small $U$, the system develops a CDW with a gap determined by the SOC, such that the density order does not gap the helical edge states [see Fig.\,\ref{SCpairing}(h)].
In the absence of SOC, this state would correspond to a gapless CDW with Dirac nodes. 
A schematic phase diagram of the strained kagome lattice is shown in Fig.\,\ref{SCpairing}(f).  
The nature of the different phases, especially contrasting those for $\epsilon=0$, Fig.\ \ref{SCpairing}(e), is indicated. Notice that the critical values of $U/t$ indicated by black stars depend on the filling factor $f$.

In summary, we have studied the tunability of electron pairing and topological states in the uniformly uniaxial strained kagome lattice with intrinsic SOC. 
The attractive Hubbard Hamiltonian in the mean-field approximation predicts that strain would create a phase where an interaction-driven CDW coexists with a topological state due to the intrinsic SOC. 
Increasing strain or interaction strength produces a topological transition to a strong CDW with a trivial gap. 
The nematicity associated with this translational-symmetry-preserving CDW becomes highly tunable by the direction of the strain field.
When superconducting pairing is allowed, the system develops a superconducting order which coexists with a CDW when strain is applied to the lattice. This superconducting order is eventually suppressed by the CDW for increasing interaction strength. 
These findings suggest that the competition between CDWs and superconductivity in kagome superconductors could be tuned by strain fields. As different experimental probes can be implemented to study such phenomena, we eagerly look forward to the verification of this behavior.

Supported by U.S. Department of Energy, Office of Basic Energy Sciences, Materials Science and Engineering Division.

\bibliography{biblio.bib}
\end{document}